\documentclass[11pt]{article}
\usepackage[utf8]{inputenc}
\usepackage{amsmath}
\usepackage{amsfonts}
\usepackage{amssymb}
\usepackage{makeidx}
\usepackage{graphicx}
\usepackage{achemso}
\usepackage{setspace}
\usepackage{pgfplots}

\usepackage[left=2.5cm,right=2.5cm,top=3cm,bottom=3cm]{geometry}
\begin{document}
\title{Recent developments in methods for identifying reaction coordinates}
\author{Wenjin Li and Ao Ma*\\
[11pt] Department of Bioengineering\\
The University of Illinois at Chicago\\
851 South Morgan St.\\
Chicago, IL 60607\\ \\
* correspondence should be addressed to:\\
Ao Ma\\
Email: aoma@uic.edu\\
Tel: (312)996-7225
}
\date{}
\maketitle

\begin{abstract}
In the study of rare events in complex systems with many degrees of freedom, a key element is to identify the reaction coordinates of a given process.  Over recent years, a number of methods and protocols have been developed to extract the reaction coordinates based on limited information from molecular dynamics simulations.  In this review, we provide a brief survey over a number of major methods developed in the past decade, some of which are discussed in greater detail, to provide an overview of the problems that are partially solved and challenges that still remain.  A particular emphasis has been placed on methods for identifying reaction coordinates that are related to the committor.   


\end{abstract}

\section{Introduction}
Many essential biological and biochemical processes, such as protein folding, conformational dynamics and enzymatic reactions, are rare events in the sense that they occur on time scales that are orders of magnitude slower than that of the elementary molecular motions.  A standard simplified picture of rare events is a transition along a special degree of freedom termed the reaction coordinate between two stable states that are separated by a free energy barrier that is high compared to the thermal energy $k_{B}T$.  This picture has its root in the transition state theory (TST)~\cite{Wigner:1938aa,Chandler:1978aa} and Kramers theory~\cite{Kramers:1940aa} for chemical reaction dynamics, in which the two stable states are the reactant and product states and the energy barrier locates the transition state.  The transition from the reactant to the product requires overcoming the high free energy barrier between them, which induces the separation in the time scale of reactive events from that of elementary atomic motions.  The rare event nature of reactive processes and the critical role of the transition state in such processes has brought the reaction coordinate to a central stage in today's computational studies of complex systems.

In the earlier developement of TST for small molecular systems, the identity of reaction coordinates was implicitly assumed as self-evident.  Then the reaction dynamics of the system is determined by the free energy profile (FEP) and diffusion coefficient along the reaction coordinate.  This situation changed dramatically when the focus of investigations on transition processes shifted to complex systems, where it was often found that the actual identities of the reaction coordinates, if happen to be known, are more than often counter-intuitive.  

More specifically, the significance of reaction coordinates to studies of reactive dynamics of complex systems is reflected in the following aspects.  First, knowledge of the correct reaction coordinates provide the fundamental details of the underlying mechanisms of a given transition process.  The FEP along the reaction coordinates  allow us to determine the activation energy and transition states, and thus the essence of the reaction dynamics.  In particular, the FEP provides a projection of the dynamics in high-dimensional space onto a few degrees of freedom that allows an intuitive and immediate grasp of a complex process.  On the more practical side, reaction coordinates are also intimately related to development of effective enhanced sampling methods.  Straightforward molecular dynamics (MD) simulations spend the vast majority of simulation time sampling stable regions, whereas the more interesting transition regions are rarely visited, if at all.  In order to study these rare events at the atomistic level, various enhance sampling methods, e.g., umbrella sampling~\cite{Torrie:1977aa}, metadynamics~\cite{Laio:2008aa}, orthogonal space sampling~\cite{Zheng:2008ab, Zheng:2009aa, Zheng:2012aa} and constrained dynamics~\cite{Sprik:1998aa, Li:2012aa}, have been developed to improve sampling of regions other than the stable basins.  These methods rely on application of a biasing potential on one or a small set of coordinates, usually termed reaction coordinates or order parameters, along which the progress of the transition can be quantified to certain extent.  In this regard, the best coordinates to apply bias are the reaction coordinates, as the bias on the correct reaction coordinates will guide the simulation through the true dynamic bottleneck in the configuration space for the given process. 

Inspite of the importance of a reaction coordinates, systematic research on how to identify them is a relatively new field and still at a rather primitive stage.  The early studies on this subject bear a ``trial-and-error'' flavor and more systematic methods have only started to be developed recently.  A few reasons contributed to this situation and two of the most prominent challenges are: 1) it is computationally demanding to obtain data from molecular dynamics simulations of complex macromolecular systems that is sufficient for the purpose of determining the reaction coordinates, and 2) given the sufficient data, correctly picking a few coordinates out the the enormous degrees of freedom of a complex system is in itself a challenging task.  

The existing methods can be categorized from two aspects.  On the one hand, they can be grouped according to the definition of reaction coordinates.  In this regards, there are mainly two different views: 1) reaction coordinates should reveal the underlying mechanism of the process under study, and 2) reaction coordinates should provide a reduced description of a given process that preserves some geometric or informatic metric of the configuration space of the system.  Free energy related definition~\cite{Krivov:2006aa, Krivov:2008aa} and committor are the prominent examples of the first group.  Committor is gaining popularity as the measure of the quality of reaction coordinates due to its clear and specific relationship with reaction dynamics.  The second category mainly includes dimensionality reduction oriented methods such as Isomap~\cite{Tenenbaum:2000aa, Das:2006aa}, diffusion map~\cite{Rohrdanz:2011aa, Coifman:2008aa, Coifman:2006aa} and Sketch-map~\cite{Ceriotti:2011aa}.  

On the other hand, existing methods can also be categorized based on the way that the reaction coordinates are determined.  Earlier methods are heavily ``trial-and-error'' in nature: a structural coordinate is selected based on chemical and/or physical intuition, then biased molecular simulations are performed along the proposal coordinate to collect necessary information, which is used to judge whether the proposed coordinate is a reaction coordinate or not.  As the collected information is specific to the selected coordinate, it can not be reused to test whether other coordinates are good reaction coordinate or not.  If the proposed coordinate is not a reaction coordinate, then other coordinates will be selected and new biased molecular simulations will be performed.  Due to the often counter-intuitive nature of the reaction coordinates, error occurs far more frequently than successes, making the ``trial-and-error'' approach too costly.  Consequently, more systematic methods that involve first preparing a database that contains information for determining the reaction coordinates and then using typically machine learning inspired methods to identify the reaction coordinates out of a pool of candidates.  Methods such as the Genetic Neural Network~\cite{Ma:2005aa} , Likelihood Maximization method~\cite{Peters:2006aa, Peters:2007aa, Peters:2012aa}, nonlinear reaction coordinate analysis~\cite{Lechner:2010aa} and Kernel PCA~\cite{Antoniou:2011aa} all belong to this group.

In this review, we discuss methods developed to identify the reaction coordinate, with emphasis on the methods that using the committor as the ideal reaction coordinate.  In addition, the history and theoretical development of the committor will be introduced, as the methods based on the committor are becoming the dominant ones in the field and the committor itself is commonly used to check the quality of a coordinate as the reaction coordinate.  We are aware of the limitations of the current review, as we mainly discussed methods that are close to our own field of expertise and interest.

\section{Committor based methods} 
In this section, we focus our discussion on methods that use the committor as the ideal reaction coordinate and judge the quality of any given physical coordinates as reaction coordinates based on their relationship with the committor.
\subsection{Committor}
For a transition between two stable basins, a trajectory initiating from a given configuration will commit to one of the basins.  The probability of an arbitrary trajectory from a configuration to commit to the product state  before the reactant state takes a fixed value for a given equilibrium ensemble, thus this probability quantifies how close a configuration is to the product state in a parametric manner.  Onsager is the first one who used the probability that two particles will not combine to quantify the progress of ion-pair recombination~\cite{Onsager:1938aa}.  This probability was termed the splitting probability and its expression in diffusive systems were derived by Kampen and Gardiner~\cite{Van-Kampen:1978aa, Gardiner:1985aa}.  Pratt and Ryter appear to be the first to define transition state using the concept of splitting probability~\cite{Pratt:1986aa, Ryter:1987aa, Ryter:1987ab}––they defined transition states as the  states with a splitting probability of 0.5, namely the states with equal probability to relax to the reactant and the product.  Therefore the splitting probability can be used to test whether a configuration is a transition state or not.  Later on, this definition has been used in the study of activated escape of a Brownian particle from a potential well~\cite{Klosek:1991aa}, protein folding~\cite{Du:1998aa}, where the name $p_{\rm fold}$ was used, and ion pair dissociation in water~\cite{Geissler:1999aa}, where the spliting probability was termed as the commitment probability or committor (committor is used consistently in the review).  Nowadays, the committor is widely used in the study of chemical and biochemical processes~\cite{Bolhuis:2000ab, Ma:2005aa, Best:2005aa, Quaytman:2007aa, Juraszek:2006aa, Hu:2008aa, Li:2010aa, Snow:2006aa, Hagan:2003aa, Radhakrishnan:2003aa, Moroni:2005aa, Gsponer:2002aa, Chu:2004aa}.

Pratt proposed to initiate a number of Monte Carlo simulations from a configuration and use the fraction of trajectories that return to a reactant configuration to define the transition state~\cite{Pratt:1986aa}.  This is equivalent to the committor, where the probability to the product region is counted.  Such a procedure was employed by Du \textit{et. al.,} to calculate the committor of protein folding~\cite{Du:1998aa}––the first practical application of committor to evaluate whether a geometrical coordinate is a good reaction coordinate or not.  Later on a similar procedure termed shooting  was proposed by Geissler \textit{et. al.,} to evaluate the committor with MD simulations~\cite{Geissler:1999aa}.  By definition, the transition states that are determined by a good reaction coordinate should  have a committor value of 0.5 or a narrow distribution of the committor value centered around 0.5.  Now this Committor Histogram Test (CHIT) that tests the reaction coordinate based on the histogram of committor values of configurations with the critical value of the proposed reaction coordinate is widely adopted in various studies~\cite{Bolhuis:2000ab, Quaytman:2007aa, Ma:2005aa, Best:2005aa, Peters:2006aa, Peters:2007aa, Chen:2009aa, Cao:2013aa}.  Recently, Peters analyzed the statistical error of CHIT in the estimation of reaction coordinate~\cite{Peters:2006ab}.

The theoretical study of the committor was rare until recent years.  For a system that can be described by the Smoluchowski equation, Gardiner derived the analytic formula of the committor~\cite{Gardiner:1985aa}.  Based on the analytic expression of committor, Rhee and Pande~\cite{Rhee:2005aa} demonstrated that committor is the reaction coordinate of a diffusive process with a parabolic barrier at the saddle point of the potential of mean force (PMF).  When the direction of the reaction is defined as the gradient along the committor, it was found to be parallel to the eigenvector of the matrix $VD$, where $V$ is the Hessian matrix of the PMF and $D$ is the diffusion coefficient matrix.  Under the same parabolic barrier approximation, Berezhkovshiii and Szabo~\cite{Berezhkovskii:2005aa} demonstrated that the PMF along the eigenvector of  the matrix $VD$ preserve the exact mean first passage times and thus the rate constant of a diffusive process predicted by the multidimensional Kramers–Langer theory.  Therefore, the PMF along the committor can reproduce the exact rate constant.  In addition, Rhee and Pande~\cite{Rhee:2005aa} have proposed a reaction coordinate whose PMF can reproduce the probability density function (PDF) of the committor for configurations in an equilibrium ensemble but may not preserve the right rate constant.  The construction of  such a coordinate requires knowledge of the PDF of the committor, which is computationally expensive, and the proposed reaction coordinate is not a physical coordinate.  Recently, a more general diffusion equation along the committor is derived by projecting multidimensional diffusive dynamics onto it, assuming the committor is the slowest coordinate of the system~\cite{Berezhkovskii:2013aa}.  It showed that the resulting diffusion equation preserves the exact reactive flux at equilibrium and thus the rate constant but not the exact dynamics.  Therefore the committor can be considered as an ``ideal'' reaction coordinate.

Remarkably, the above-mentioned studies~\cite{Berezhkovskii:2005aa, Rhee:2005aa} showed that the reaction coordinate, which is perpendicular to the isocommittor surface, is not necessarily parallel to the gradient of the PMF at the saddle point (the eigenvector of the matrix $V$) for anisotropic diffusion systems.  It is actually parallel to the eigenvector of the matrix $VD$.  Ma \textit{et. al.,} have verified these results based on a study of an isomerization reaction of an alanine dipeptide in implicit solvent and found that the two eigenvectors deviate by a small angle~\cite{Ma:2006aa}.  The PMF and the diffusion tensor along two reaction coordinates were estimated from MD simulations, and then committor were estimated based on Berezhkovshiii and Szabo's theory~\cite{Berezhkovskii:2005aa}.  The predicted committor is consistent with the one estimated with MD simulations, demonstrating that a complex biological process can be simplified by a low-dimensional physical model.

\subsection{``Trial-and-error" methods}
Most early work on the searching of reaction coordinate employed a trial-and-error process.  Examples include but are not limited to the studies on simple solvated systems~\cite{Geissler:1999aa}, enzymatic reactions~\cite{Quaytman:2007aa, Li:2012aa, Klahn:2006aa, Rosta:2009aa}, protein folding~\cite{Dinner:1999aa, Du:1998aa}, and biomolecular conformational changes~\cite{Hagan:2003aa, Bolhuis:2000ab}.  Typically a reaction coordinate is proposed based on intuition and knowledge, and information is then collected to test whether it is a reaction coordinate or not based on various mechanism oriented criteria.  For instance, free energy along the proposed coordinate was estimated with enhanced sampling methods and the reaction coordinate was considered to be the one with highest free energy barrier~\cite{Rosta:2009aa} or the reaction coordinate can be approximated by a minimum free energy path with a free energy barrier consistent with experimental results~\cite{Li:2012aa, Klahn:2006aa}.  More rigorous criteria are based on the committor, where committors of selected configurations at fix values of a coordinate were estimated and CHIT was commonly used to determine the quality of the coordinate as a reaction coordinate~\cite{Du:1998aa, Bolhuis:2000ab, Geissler:1999aa, Hagan:2003aa, Quaytman:2007aa}.  

\subsection{Methods based on $p({\rm TP}|r)$}

Best and Hummer~\cite{Hummer:2004aa, Best:2005aa} proposed that transition states are configurations with the highest $p({\rm TP}|x)$, where $p({\rm TP}|x)$ is the probability for a trajectory that passes through a configuration $x$ to be a transition path.  And the reaction coordinate is a coordinate $r$ with sharpest peak in $p({\rm TP}|r)$, where $p({\rm TP}|r)$ is defined as $p({\rm TP}|r)=\frac{\int p({\rm TP}|x) \delta[r-r(x)]p_{\rm \,eq}(r)dx}{\int \delta[r-r(x)]p_{\rm \,eq}(r)dx}$, which is the average probability for trajectories passing through configurations with the same value of $r$ to be transition paths.  Here, $\delta(r)$ is the Dirac's delta function and $p_{\rm \,eq}(r)$ is the equilibrium probability distribution of system configurations projected onto the coordinate $r$.  Therefore any coordinate can be tested by obtaining $p({\rm TP}|r)$ from a long equilibrium trajectory and the coordinate with the highest peak of $p({\rm TP}|r)$ can therefore be identified as the reaction coordinate.  Since it is not always possible to prepare a long enough equilibrium trajectory, they proposed a computationally less costly way to estimate $p({\rm TP}|r)$.  According to a Bayesian relationship between the equilibrium ensemble and the transition path ensemble, $p({\rm TP}|r)=p(r|{\rm TP})p({\rm TP})/p_{\rm \,eq}(r)$ for Markovian processes.  Here, $p_{\rm \,eq}(r)$ and $p(r|{\rm TP})$ are the probability distribution of configurations projected onto the coordinate $r$ for the equilibrium ensemble and the transition path ensemble, respectively.  $p({\rm TP})$ is the fraction of time that the system spent in the transition paths, relative to the total time in the long equilibrium trajectory.  $p_{\rm \,eq}(r)$ of a chosen coordinate can be obtained by enhanced sampling methods, e.g., umbrella sampling was used in the work by Best and Hummer~\cite{Best:2005aa}, to reduce the computational cost.  The optimization of $p({\rm TP}|r)$ was later demonstrated to be equivalent to a method that optimize the stochastic separatrix~\cite{Peters:2010aa}.

\subsection{Methods that utilize machine learning algorithms}
The biggest advantage of collecting sufficient data first and analyzing them to search for the reaction coordinate is that one can consider every possible coordinate as candidate of reaction coordinate without assuming that the reaction coordinates are contained in a small set of pre-selected collective variables.  This assumption is not necessarily true, as the identities of the correct reaction coordinates are often counter-intuitive.  What is the sufficient information to identify the reaction coordinate?  A long equilibrium trajectory that samples enough transitions between stable states should be sufficient, although the preparation of such a trajectory is only possible for small systems with MD simulations or a few medium systems by highly parallel distributed computing or by special purpose high-performance computing~\cite{Snow:2002aa, Lane:2011aa, Bowman:2010aa, Piana:2012aa}.  Since committor is the ideal reaction coordinate, committor of configurations in the transition path ensemble should contain sufficient information to extract the reaction coordinate~\cite{Ma:2005aa, Peters:2006aa, Peters:2007aa, Lechner:2010aa}.  In this regard, transition path sampling (TPS)~\cite{Dellago:1998aa, Dellago:1999aa, Bolhuis:1998aa, Bolhuis:2002aa} and committor estimation are the most commonly used way to harvest the information. 

\paragraph{Genetic Neural Network Method} The method developed by Ma and Dinner~\cite{Ma:2005aa} uses a genetic neural network (GNN)~\cite{So:1996aa, So:1996ab} method to extract the reaction coordinates from a pool of pre-selected cndidates based on the committor information contained in a database of systems configurations.  The committor value for each configuration in the database was accurately evaluated using the shooting procedure.  Configurations were harvested with transition path sampling~\cite{Dellago:1998aa, Dellago:1999aa, Bolhuis:1998aa, Bolhuis:2002aa} and selected to ensure uniform distribution of the committor values in the database to avoid bias in the training and testing of the neural network model.  The GNN method was then applied to identify the combination of coordinates that produces the most accurate prediction of the committor, which is considered the best approximation of the reaction coordinate.  

The GNN method is a combination of a genetic algorithm and a neural network method.  The neural network was used to build the model that can produce the best prediction for committor value for a given combination of physical coordinates and the genetic algorithm identifies the best combination of coordinates among all of those sampled from the candidate pool through a Monte Carlo like procedure.  The best combination of physical coordinates from the GNN procedure provides us the components of the reaction coordinates.  In addition, the GNN method provides a mathematical expression that can be used to predict the committor value of a given configuration with great accuracy.  Neural network model establishes a relationship between physical coordinates and committor, which is flexible and in principle could take into account the potentially complex nonlinear relationship between the physical coordinates and committor.  The GNN method is not intrinsically coupled to transition path sampling,  other methods for harvesting transition paths, such as transition interface sampling~\cite{Erp:2003aa, Van-Erp:2005aa} and forward flux sampling~\cite{Allen:2009aa}, should work as well. 

The GNN method has been applied to alanine dipeptide isomerization reactions, a standard model system for reaction coordinate studies, in vacuum and implicit and explicit water.  In the vacuum, the identified reaction coordinates are consistent with the results of a previous study~\cite{Bolhuis:2000ab}.  In explicit water, a torque on the solute derived from electrostatic interactions with the solvent molecules was found to be a critical component of the reaction coordinate.  CHIT showed that the identified reaction coordinate can predict transition states with committor values narrowly distributed around 0.5.  In implicit water, the predicted components of the reaction coordinate were successfully used to construct a low-dimensional physical model to describe the dynamics of the complex biomolecular system~\cite{Ma:2006aa}.  Other applications of the method included the study of a nucleotide flipping facilitated by O$^{\rm 6}$-alkylguanine-DNA alkyltransferase~\cite{Hu:2008aa} and the study of the folding of a 20-residue antiparallel-sheet miniprotein~\cite{Qi:2010aa}.

\paragraph{Likelihood Maximization Method} Peters \textit{et. al.,}~\cite{Peters:2006aa} have designed an aimless shooting algorithm for TPS, in which the momenta of the system for a given configuration are drawn from the Boltzmann distribution, instead of being derived from a small perturbation of the original momenta as implemented in the original TPS algorithm.  In aimless shooting, each shooting trajectory can be considered as a realization of committor and a large set of configurations with committor estimated by such a one-time realization can be collected from the TPS history.  For a given set of coordinates, the relationship between the committor and a linear combination of coordinates is proposed to be a sigmoid function with $P_{\rm B}(r)=[1+{\rm tanh}(r)]/2$, where $P_{\rm B}(r)$ is the averaged committor of configurations along the reaction coordinate $r$ and $r$ is a linear combination of several physical coordinates.  The reaction coordinate is approximated by the linear combination that maximizes the likelihood~\cite{Peters:2007aa}:

\begin{equation}
L=\prod_{P_{\rm B}(x_{\rm k})=1} P_{\rm B}(r(x_{\rm k}))\prod_{P_{\rm B}(x_{\rm k})=0} (1-P_{\rm B}(r(x_{\rm k})))
\end{equation}

It is the occurring probability of the one-time committor realization of these configurations assuming that the committor along the reaction coordinate $r$ is $P_{\rm B}(r)$.  Here $x_{\rm k}$ is one of the configurations whose committor is estimated, $P_{\rm B}(x_{\rm k})$ is the estimated committor of a configuration $x_{\rm k}$ by a single shooting trajectory, and $P_{\rm B}(r(x_{\rm k}))$ is the committor of $x_{\rm k}$ estimated by the proposed sigmoid function. 

Different numbers of coordinates and different combinations of coordinates can be tested to find the best approximation of the reaction coordinate by taking the combination of coordinates with the maximum likelihood.  Typically, the more coordinates are included, the higher likelihood for the resulting model.  The optimal number of coordinates is reached if there is no significant increase of the likelihood when an extra coordinate is taken into account~\cite{Peters:2006aa}.  The distribution of the configurations in the database was assumed to be peaked near the transition state region, as the aimless shooting procedure has the tendency to concentrate towards the transition state.  A recent extension of the likelihood maximization is the inertial likelihood maximization method~\cite{Peters:2012aa}, which takes into account the velocities projected onto the selected coordinates as well.  For the systems studied by this method, the variance of the committor values of configurations on the transition state surface that is determined by the optimized reaction coordinate is in general smaller than the ones obtained by the original likelihood maximization method.  In addition, the transmission coefficients of proposed transition states from inertial likelihood maximization are larger and closer to 1.  Thus the inertial likelihood maximization method is an improvement over the original likelihood maximization approach.  Recently, Lechner \textit{et. al.,} introduced non-linearity into the reaction coordinate in the likelihood maximization method~\cite{Lechner:2010aa}.  Learning from the string methods~\cite{Weinan:2002aa, Weinan:2007aa, Maragliano:2006aa, Maragliano:2007aa, Weinan:2006aa}, they replaced the linear combination of coordinates by a string of configurations in a low-dimensional collective variable space to approximate the reaction coordinate using the likelihood maximization and the committor of configurations was obtained from a replica exchange transition interface sampling.

The likelihood maximization method have been applied to a number of systems: the mechanism of the partial unfolding transition in a photoactive yellow protein~\cite{Vreede:2010aa}; the folding details of Trp-cage protein in explicit solvent~\cite{Juraszek:2008aa}; the homogeneous nucleation process of a crystal in a Gaussian core model~\cite{Lechner:2011ab, Lechner:2011aa}; diffusion of water molecules in a glassy polymer~\cite{Xi:2013aa}.

The likelihood maximization approach shares a number of similarities with the GNN method discussed earlier.  Both methods assume the committor of configurations as the sufficient information for identifying the reaction coordinate.  In the former, the committor is evaluated in great accuracy, whereas in the latter the committor is estimated by a one-time realization.  In the GNN method, the distribution of configurations along the committor is enforced to be uniform, whereas it is a natural outcome of the aimless shooting procedure in the likelihood maximization method , which will in principle vary with the system under study but is likely to concentrate around the committor value of 0.5 due to the particular feature of aimless shooting.  To extract the reaction coordinate from the given information, both methods resort to a sigmoid model.  In the GNN method, the sigmoid model is employed inside the neural network, whereas in the likelihood maximization method, it directly establishes the relationship between the committor and the coordinates.  In fact, the sigmoid model in the likelihood maximization can be considered as a specific neural network model, in which there is no hidden layers––with the selected coordinates as input and the committor as the only output.  Also one can hybrid the two methods together.  For instance, the likelihood maximization procedure can be applied to a set of configurations with committor values estimated from a standard shooting procedure instead of the aimless shooting, although such a hybrid method was not able to identify the reaction coordinate in a study to the thiol/disulfide exchange in a protein~\cite{Li:2010aa}. 

\paragraph{Transition state ensemble optimization}
In the standard picture of reaction dynamics, the reaction coordinate is perpendicular to the transition state surface at the saddle point of the free energy surface, methods were thus developed to identify geometrical coordinates that comprise a reaction coordinate in complex systems from the transition state ensemble alone.  The transition state ensemble or the stochastic separatrix is a hypersurface in the configuration space on which there is no change in the progression of the reaction~\cite{Antoniou:2009aa}.  To characterize the behavior of coordinates on the stochastic separatrix, Antoniou and Schwartz~\cite{Antoniou:2009aa} plotted the distributions of all the coordinates and examined the width of their distributions on the separatrix in model systems of a double well potential coupled to a single and multiple oscillators.  They found that the variation of the reaction coordinate is significantly smaller than that of the non-reactive coordinates.  Based on this observation, they proposed a method to identify coordinates with the smallest variations along the separatrix as the reaction coordinate.  The method assumes a narrow transition state region and is therefore mainly applicable to systems with fast barrier crossing, but not so much to systems with diffusive barrier crossing~\cite{Antoniou:2009aa, Antoniou:2011ab}. 

In a later application, this method was generalized to deal with high dimensionality and nonlinearity that are often present in real systems.  Their basic assumption is that the width of the separatrix along a coordinate that is part of the reaction coordinate must be thin~\citep{Antoniou:2011aa}, as the correct reaction coordinate should take a fixed value at the separatrix.  On the other hand, the non-reaction coordinates will show large variations in the transition state ensemble as they do not correlate with the progression of the reaction.  Therefore one can check the width of the separatrix along all the coordinates and select those with narrow width as the components of the reaction coordinate.  Mathematically, the width of the separatrix along a coordinate can be quantified by the contribution of the coordinate to the direction of maximum variance on the separatrix.  Coordinates that contribute little to the direction of maximum variance will be part of the reaction coordinate.  Traditionally, principal component analysis is the standard method for identifying the direction of maximum variance.  However, the separatrix of a complex system is a curved surface with great complexity and it is non-trial to identify the dominant direction of the separatrix.  Consequently they proposed a kernel principal component analysis (kPCA) approach with a non-linear kernel function to identify the direction with the largest variance.  The method has been tested on an enzymatic reaction catalyzed by lactate dehydrogenase, which is previously studied by the same group in great details with TPS and committor analysis~\cite{Quaytman:2007aa, Basner:2005aa}.  The identified reaction coordinates by kPCA are consistent with those previously obtained with the ``trial-and-error'' method~\cite{Quaytman:2007aa}. 

\section{Free energy related methods}
Minimum energy path and minimum free energy path are conventionally associated with the concept of reaction coordinate~\cite{Fukui:1970aa, Quapp:1984aa, Gonzalez:1990aa} and this connection has recently been derived from transition path theory as well~\cite{Vanden-Eijnden:2006aa, Weinan:2006aa, Weinan:2010aa}.  For simple chemical reactions, the minimum energy path can be found by searching along all the degrees of freedom.  However, this procedure is practically impossible for complex systems due to the large number of degrees of freedom.  Consequently, searching over several collective variables is employed for macromolecular systems and the minimum free energy path is searched for instead.  

Various methods have been developed to find the minimum free energy paths or maximum flux paths for complex systems, e.g., string methods~\cite{Weinan:2002aa, Weinan:2007aa, Maragliano:2006aa, Maragliano:2007aa, Weinan:2006aa} and elastic band methods~\cite{Henkelman:2000aa, Sheppard:2008aa, Jonsson:1998aa}. Both types of methods start with a chain of states that connects the reactant and the product, then this chain of states are evolved towards the minimum free energy paths using optimization algorithms. In string methods the evolution of different states are essentially independent of each other, whereas in elastic band methods each state is subject to constraints from neighboring states. Since the conformational searching is performed in a low-dimensional space spanned by a few collective variables, the choice of the collective variables is quite crucial for the success of these methods. These methods and their extended ones have been successfully applied to studies of biological and material systems~\cite{Elder:2012aa, Miller:2007aa, Ren:2005ab, Yazyev:2008aa, Greeley:2004aa, Krasheninnikov:2009aa, Chen:2009aa, Cao:2013aa}. For more details of these methods, we refer readers to recent reviews~\cite{Weinan:2010aa, Rohrdanz:2013aa}.

Krivov and Karplus proposed the concept of the cut-based free energy profile~\cite{Krivov:2006aa, Krivov:2008aa}, which was inspired by their earlier work on free energy disconnectivity map~\cite{Krivov:2002aa, Krivov:2004aa}.  Previously, they showed that the partition function of the reactant region can preserve the energy barriers~\cite{Krivov:2006aa, Krivov:2008aa}.  For conventional free energy profile, its partition function ($Z_{\rm H}$) is constructed from the density of configurations along a given coordinate.  For cut-based free energy profile, its partition function ($Z_{\rm C}$) is proportional to the total number of transitions through a given point on the coordinate during a small time interval.  There exists a useful relationship between $Z_{\rm C}$ and $Z_{\rm H}$, in which $Z_{\rm C}$ can be expressed as a function of $Z_{\rm H}$ and position-dependent diffusion coefficient.  Most importantly, a ``natural coordinate'', which is a continuous and invertible function of a given coordinate, can be constructed such that $Z_{\rm C}$ and $Z_{\rm H}$ along the natural coordinate differ by a constant or the diffusion coefficient along the natural coordinate coordinate is constant.  They demonstrated that the reaction coordinate is the one with the highest cut-based free energy barrier along its natural coordinate~\cite{Krivov:2008aa}.  

Later, Krivov proposed the idea of the cut free energy profile, whose partition function is proportional to the sum of the diffusion distance from a give point during a small time interval, and the reaction coordinate is the one with a constant cut free energy profile, which is position and sampling interval independent~\cite{Krivov:2012aa}.  A major concern of the above-mentioned methods is that it is almost computational impossible to get a long enough equilibrium trajectory for real systems and it is not trivial to construct and compare two cut-based free energy profiles.

\section{Dimensionality reduction oriented methods}
Dimensionality reduction, which means projecting a high dimensional data set onto a few essential directions to gain a clear visualization and facilitate conceptualization of otherwise extremely messy and complex data, is an important subject in the fields of informatics and data mining in general.  The same philosophy has been utilized in the studies of protein dynamics, with principal component analysis~\cite{Jolliffe:2005aa} as the most familiar example.  Recently, some of the most popular methods for general dimensionality reduction purpose, such as Isomap~\cite{Tenenbaum:2000aa, Das:2006aa}, diffusion map~\cite{Rohrdanz:2011aa, Coifman:2008aa, Coifman:2006aa, Coifman:2005aa} and sketch-map~\cite{Ceriotti:2011aa}, have been introduced to studies of reaction coordinates.  These methods typically achieve dimensionality reduction based on preservation of geometric measures of the  configuration space and  the dynamical information contained in the temporal sequence of the configurations is usually ignored~\cite{Krivov:2011aa}.  Here we briefly discuss a few of them.  

In Isomap, the separation of two configurations in the configuration space is quantified by the geodesic distance, which is defined as the shortest path between them in a connected graph~\cite{Tenenbaum:2000aa}.  A configuration is connected to another configuration if the root mean square deviation (RMSD) is small or they are nearest neighbours.  The underlying assumption is that configurations with small RMSD are mutually accessible without crossing any barrier, \textit{i.e.}, they belong to the same state, although there are configurations with small RMSD that are separated by high energy barrier.  A study of protein folding with a coarse-grained model showed that the transition state ensemble can be correctly identified from a free energy landscape in a low-dimensional space, which is constructed with Isomap~\cite{Das:2006aa}.  

Diffusion map is designed to perform non-linear dimensionality reduction of a connected graph of points.  From MD simulation data, connected graph of configurations is constructed with certain metric or kernel which quantifies the weight between two configurations.  A Markov process on the graph is then constructed by renormalizing the kernel so that it quantifies the probability for a random walker on the graph to make a step from one point to another.  The diffusion distance quantifies the rate of connectivity of two points in the graph and is robust to perturbation on the data and preserved in the dimensionality reduction~\cite{Coifman:2005aa}.  Dimensionality reduction can be performed to construct a few principal diffusion coordinate, whose ordinary Euclidean distance in the embedding space measures the diffusion distance.  Then the diffusion coordinates are correlated with quantities with more specific physical meaning~\cite{Rohrdanz:2013aa}.  For more details of these methods, we recommend an excellent recent review~\cite{Rohrdanz:2013aa}.

\section{Conclusions and Prospectives}
As our discussions suggested, the notion of reaction coordinates plays an essential role in today's understanding of the reaction mechanism of complex systems and development of efficient methods for simulating such systems.  Consequently, considerable efforts have been devoted to developing methods for identifying reaction coordinates in complex systems over the past decade.  Based on the definition of the reaction coordinates, these methods can be roughly grouped as those targeted at committor based reaction coordinates and those targeted for reduced description of a reaction process that can preserve certain geometric measures in the configuration space.  While the committor based reaction coordinates have clear mechanistic implications to the systems and processes under study, the specific meaning of configuration space geometry related reaction coordinates may require our improved understanding of the geometric picture of reactive dynamics.  

The emphasis of this review has been on the methods for identifying committor-based reaction coordinates and the current trend along this direction has been along the line of using machine learning methods to extract coordinates that exhibit strong "patterns", defined by various committer-based metrics, from a candidate pool in a database prepared from simulation data.  Along this direction, there are three main issues: 1). the sufficient information required to correctly determine the reaction coordinates from the database, 2) the computational cost involved in preparing the database, 3) principles for constructing a complete candidate pool in the sense that the inclusion of the correct reaction coordinates is guaranteed.  

For the first question, the GNN method~\cite{Ma:2005aa} requires a database that contains accurate committor information over the entire range from 0 to 1, whereas the kernel PCA method~\cite{Antoniou:2009aa, Antoniou:2011aa} only requires information around the committor value of 0.5, which reduced the computational cost considerably but may also limit the applicability of the method to reactions with high barrier and ballistic dynamics.  The likelyhood maximization method reduced the computational cost at the expense of the accuracy of committor information in the database, which could potentially limit the reliability of the reaction coordinates selected by the method.  Therefore, how to properly balance between the accuracy and range of committor information and the computational cost remains a challenge at the current stage, whereas an improved understanding of the minimum sufficient information required for reliable selection of reaction coordinates will greatly help to clarify the situation.  On the other hand, the current approach for preparing the candidate pool is intuition based and non-systematic.  A drawback of such an approach is it is difficult to determine when the candidate pool should be complete.  A clear example in this regard is the solvent coordinate essential for the isomerization reaction of the alanine dipeptide in explicit water~\cite{Ma:2005aa}, which turned out to be the electrostatic torque on the solute from the solvent molecules, a highly counter-intuitive coordinate that one would be expected to consider only after sufficient amount of trial and errors.  Of course, our intuition on this would improve after more complex systems have been studied using the rigorous approach, but a more first-principle based systematic approach should be the direction to pursue.

\providecommand{\latin}[1]{#1}
\providecommand*\mcitethebibliography{\thebibliography}
\csname @ifundefined\endcsname{endmcitethebibliography}
  {\let\endmcitethebibliography\endthebibliography}{}

  
\end{document}